\documentclass[12pt,letterpaper]{article}
\usepackage{amsmath}
\usepackage{amssymb}
\usepackage[pdftex]{graphicx}
\usepackage{latexsym}
\usepackage{graphicx}
\usepackage{color}
\usepackage{hyperref}

\def\be{\begin{equation}}
\def\ee{\end{equation}}

\def\lbldef#1#2{\expandafter\gdef\csname #1\endcsname {#2}}

\begin{document}
\baselineskip=16pt \pagestyle{plain} \setcounter{page}{1}

%Title page

\begin{titlepage}

\begin{flushright}
%hep-th/yymmnnn \\
TAUP-2856/07
\end{flushright}
\vfil

\begin{center}
{\Large\bf On Black Hole Remnants}
\end{center}

%\vfil
\begin{center}
%{\large
{\sc Aharon Casher} \footnote{e-mail: {\tt ronyc@post.tau.ac.il}}
and {\sc Nir Raz}
\footnote{e-mail: {\tt raznir@post.tau.ac.il}}\\
 %\vspace*{1cm}

{\it  {Raymond and Beverly Sackler Faculty of Exact Sciences\\
School of Physics and Astronomy\\
Tel-Aviv University , Ramat-Aviv 69978, Israel}} \\

\end{center}

\begin{center}
\begin{abstract}
 We introduce two models for a planck scale black hole
remnant (Planckon), which can hold arbitrarily large information,
while keeping a vanishing coupling and discuss their physical
properties.
\end{abstract}
\end{center}

%\noindent

\vfil
\begin{flushleft}

May, 2007
\end{flushleft}
\end{titlepage}
\newpage

\renewcommand{\baselinestretch}{1.2}
\setcounter{footnote}{0}

\tableofcontents
%\newpage

\section{Introduction}
\label{sec:intro}

\subsection{The Black-Hole Information Paradox}
\label{subsec:paradox}

Hawking's semi-classical calculation of black holes radiance [1]
led to the possibility of black hole evaporation. This in turn,
led to the conclusion that information may be lost forever, in
practice, from the world - this is the black-hole information
paradox.

The paradox may be described in the following manner:

In 1975 Hawking calculated the emission of radiation from a
stationary classical black hole. The calculation was done using a
semi-classical calculation for non-interacting matter fields
propagating over classical Schwarzschild black hole geometry.

This calculation resulted in a surprising discovery that a black
hole not only radiates, but radiates as a black-body with
temperature of:

\be T_{BH} = \frac{\hbar c^3 \kappa }{2\pi k_B G} \ee

Where $\kappa $ is the surface gravity (for a Schwarzschild black
hole $\kappa = \frac{1}{4M})$.

Since the emitted radiation is a black-body radiation it is
exactly thermal (uncorrelated), in particular, the emitted
radiation does not depend on the structure of the collapsed body
that formed the radiating black-hole (the collapsed body is
trapped behind the horizon and is unable to influence anything
outside the horizon).

The radiation depends only on the geometry of the black hole
outside the horizon (depends only of the mass, angular momentum,
charge etc. of the black hole) and can not depend or be correlated
with the collapsed body (there might be some weak correlations
since Hawking's calculation is not exact).

By itself, the fact that the radiation outside the black hole is
thermal is not too disturbing, since it is only a part of a whole
quantum system. Part of the quantum system is inaccessible, as it
is trapped behind the horizon. There are some correlations between
the degrees of freedom, which are accessible outside the horizon
and the ones inside the horizon. Because of the correlations, an
observer outside the horizon, detecting the quantum fields
(degrees of freedom which were radiated), will not be able to
determine the exact initial quantum state of the collapsed body
and will only detect a mixed state.

During the radiation process the black hole radiates its energy
(mass) away and so, if one waits long enough, the black hole will
evaporate completely, leaving behind only the thermal radiation.
The thermal radiation, which is a mixed state, is now the whole
system. The consequences for such a process are that beginning
with a pure state does not allow one to predict with certainty,
what will the final quantum state be (the final system is a mixed
state so one can only assign probabilities to different final
states).

The evaporation of a black hole, as described above, results in a
paradox. According to the laws of quantum mechanics and field
theory - if one completely specifies the initial state of a system
and knows all the stages of its evolution, one knows the final
state of the system at all future times (this is the unitarity
postulate of quantum mechanics, which states that a system in a
pure state will stay in a pure state).

This paradox is known as "The Information Loss Paradox", since
from an initially pure state, which has zero entropy, one ends up
with a mixed state, which has non-vanishing entropy. Such a
process where information is lost indicates a non-unitary
evolution, which contradicts the laws of quantum physics.

Hawking showed, that the emitted thermal (or nearly thermal)
radiation from the evaporating black hole, carries a huge amount
of entropy that can be estimated by:

\be S \sim \frac{M_0^2 }{M_{pl}^2 } \ee

Such an evolution of a black hole, from a pure state into a mixed
state, results in a fundamental loss of information:

\be \Delta I = - \Delta S \sim - \frac{M_0^2 }{M_{pl}^2 } \ee

The source of this missing information is the correlation between
particles coming out of the black hole and particles falling into
the black hole.

The semi-classical calculation is valid until the black hole
reaches the Planck scale, where quantum gravity effects that break
the semi-classical approximation, may affect the process. The
Planck scale, which is given by Planck's mass, time and length,
can be formed by combining the gravitational constant G, the
quantum of action $\hbar $ and the speed of light c in a unique
way.

The Planck units are:

\be
\begin{array}{l}
 M_{pl} = \sqrt {\frac{\hbar c}{G}} \approx 1.22 \cdot 10^{19}GeV \approx
2.17 \cdot 10^{ - 5}[gr] \\
 t_{pl} = \sqrt {\frac{\hbar G}{c^5}} \approx 5.4 \cdot 10^{ - 44}[\sec ] \\
 l_{pl} = \sqrt {\frac{\hbar G}{c^3}} \approx 1.62 \cdot 10^{ - 33}[cm] \\
 \end{array}
\ee

Since quantum gravity is expected to play a key role in the
evaporation process, one may hope, that a resolution of the black
hole information paradox may be found as the black hole (mass and
length) approaches the planck scale by some , yet unknown, quantum
gravitational effect.

\subsection{Possible Solutions To The "Black Hole Information Paradox"}
\label{subsec:solutions}

Hawking's semi-classical calculation indicates the breakdown of
predictability and unitarity in physics. Yet the calculation is
valid only at length scales larger than the Planck scale. One
might hope that the solution to the apparent paradox will appear
at the Planck scale and thus will produce some clues, as to how to
construct a quantum gravity theory or at least shed light on some
of its features.

There are three main approaches to find a solution to the black
hole information paradox: \\
The first accepts the information loss. The second asserts that
the information is retrieved during the evaporation process or via
effects, which occur around the Planck scale. The last solution
relies on the possible existence of Planck scale remnants.

The next few paragraphs will contain a brief discussion on the
first two approaches and the rest will focus on the last solution,
which is the basis of this article.

\underline {Information loss}:

This solution tries to implement information loss into physics and
especially into quantum physics. The primary attempts are to
change or generalize the unitarity postulate of quantum mechanics
to allow non-unitary evolution. An example for such an extension
of quantum mechanics was offered by Hawking \cite{hawk2}, who
suggested replacing the usual S matrix of quantum mechanics (which
maps a pure state only to another pure state) with a
super-scattering matrix ${S}$, which acts on density matrix
(instead of on state vectors) in the following way:

\be \rho _{ab}^{final} = {S}_{ab}^{cd} \rho _{cd}^{initial} \ee

The super-scattering operator ${S}$ acts on a density matrix and
maps it into another density matrix in a non-unitary way and thus
can increase the entropy. In particular, the operator ${S}$ can
act on a pure state and map it into a density matrix (mixed
state).

The main drawbacks of accepting information loss as an unavoidable
feature of quantum gravity are that no one has yet found a way to
incorporate non-unitarity into consistent physical theories that
gives satisfactory results. Furthermore, the fact that
transmitting information requires energy indicates that losing
information might be connected to violation of energy and momentum
conservation.

Let us assume that a black hole forms and than evaporates in a
time $\Delta t$ - then from the uncertainty principle one has

\be \Delta E \geq \frac{1}{\Delta t} \ee

This shows that the minimum of energy loss in the process should
be of order $\frac{1}{\Delta t}$.

In quantum theory such formation and evaporation should occur all
the time as virtual processes. The amplitude for such processes
approaches unity when the size of the loop approaches the Planck
scale (there is no smaller dimensionless number to suppress it).
Thus, one would expect Planck size energy violations with
characteristic time of the order of Planck time. This would cause
the world to seem as a thermal bath of Planck temperature, which
is obviously not the case.

\underline {Information retrieval}:

This line of thought suggests that unitarity is not violated,
usually by suggesting that the information about the state of the
collapsing matter is encoded in the emitted Hawking radiation.
Alternatively, by suggesting that the information comes out in a
final burst when the black hole reaches the Planck scale (Planck
mass).

Both suggestions have their drawbacks:

The first suggestion implies that matter behind the horizon
affects matter outside the horizon, despite the fact that the two
regions of space-time are space-like separated. Hence, one has to
give up the notion of locality and causality or at least change
them radically (this idea also contradicts Hawking's calculation
\cite{hawk1}).

The second suggestion raises problems with energy conservation
since one can show that the information does not have enough
energy to exit in a final burst:

The energy of the black hole in the Planck scale is $E_{pl} \sim
M_{pl} $ and the information to be transmitted is of the order of
$\frac{M_0^2 }{M_{pl}^2 }$. Because of the uncertainty principle,
the only way to transmit a lot of information with little energy
is to transmit the information over a long period of time $\Delta
t$.

An estimate of the time required for the transmission given Planck
energy is:

\be t \sim \left( {\frac{M_0 }{M_{pl} }} \right)^4t_{pl} \ee

This time exceeds the age of the universe for most black holes and
thus, one is drawn to the possibility of stable or long-lived
remnants of Planck mass.

\subsection{The Remnant Solution And Its Drawbacks}
\label{subsec:remnant}

Another possibility is to assume that when the black hole reaches
the Planck scale, it tunnels due to quantum effects into a stable
or nearly-stable particle which keeps the information about the
initial state.

As was implied above - the information can not come out at the end
of the evaporation with a final burst:

For a black hole with mass M, the emitted radiation state must
contain energy M inside a sphere, whose radius is comparable to
the hawking evaporation time of the black hole $t_{Hawking} \sim
M^3$ (Working in the natural units in which the Plank mass $M_{pl}
= 1)$.

The remnant can decay into $N\simeq \left(\frac {M_0}
{M_{pl}}\right) ^2 $ quanta \cite{acn}, but such a decay is highly
suppressed because of the tiny wave function overlap factor. The
reason for this small overlap is, that the only way to transmit
all the information with small available energy, is to use very
low energy (corresponding to very long wavelength) states. The
overlap between the states wave-function and the remnant
wave-function is very
small.\\
To quantify the argument above let us examine the average
wavelength of the emitted quanta \cite{acn}:\\
The average wavelength of the final N emitted quanta is \be
\lambda \approx \left( \frac{M_{pl}}{N}\right)^{-1} \approx
NR_{pl} \ee one can easily see that the wavelength of the emitted
wave is larger by a factor of N than the size of the decaying
system (Planck size black hole has a radius of $\sim R_{pl}$).\\
The "wave-function overlap" between each of the emitted quanta and
the decaying system is therefore
$f=\frac{R^{3}_{pl}}{\lambda^3}\approx N^{-3}$. The simultaneous
emission of N quanta is supressed by a tiny factor of $f^{N}
\approx N^{-3N}$. This tiny factor will render a planck scale
remnant practically stable.

%%Let us assume that the emitted thermal radiation behaves as an
%%ideal gas and for simplicity let us assume that we are working in
%%one dimension:

%%The entropy S of a one-dimensional ideal gas with energy E and
%%length L is $S^2 \sim EL$ therefore the entropy of the black hole
%%will be $S^2 \sim M^2$ ($E \sim M;L \sim M^3)$.

%%We now have to find out the length (volume) the gas occupies when
%%it posses the same entropy S but the energy is $E \sim 1$ (Planck
%%mass). From the formulas above we get that the gas has to expand
%%by a factor of M into $L \sim M^4$.

%%If it takes a time t for the information to come out the lower
%%bound for this time t is $t \ge M^4$ (the radiation occupies a
%%sphere with radius $L \sim M^4)$ and the information can not come
%%out in a final burst.

The remnant idea comes as another way to retain the unitarity
postulate of quantum mechanics and avoid the final burst of
information by leaving behind a long lived remnant.

Considering the suppression factor above, estimates of the remnant
evaporation time $t_{remnant} $ \cite{preskill,lifetime} yield a
lower bound for $t_{remnant} $ :

\be t_{remnant} \ge \left( {\frac{M_0 }{M_{pl} }} \right)^4t_{pl}
\ee

In the last formula the Planck factors were reinstated.\\
 The long evaporation time can be understood as the decay time required for
a very long wavelength mode. One should notice that $t_{remnant}
>$ "age of the universe" which validates the claim that the planckon
is virtually stable.

One can now see that in order to retain the unitarity postulate
one has to assume a stable black hole remnant which should have
mass that is equal or near the Planck mass.

Black hole Planck scale remnants were given the name Planckons
\cite{acn}, a name that will be used from now on.

One can also ask whether a Planckon can be charged (either
electric, weak, color, etc.) or have angular momentum, since the
evaporating black hole can be spinning and/or charged (Kerr black
hole). The Hawking radiation of a Kerr black hole is such that the
black hole emits its angular momentum and charge by radiating its
charge and angular momentum away, creating charged or spinning
particles. Thus, when the black hole reaches the Planck scale, one
is left with a Schwarzschild black hole \footnote{ Since the rate
of such a process for charged black hole is proportional to some
power of $\alpha$ the time for a black hole to loss its charge
(this is true for any quantum charge) is much smaller than the
time for the black hole to reach the Planck scale} \cite{charge
loss}. Even if one does end up with a charged or spinning
Planckon, the Planckon will lose its charge or angular momentum by
pair creation of particles and "swallowing" particles with
opposite sign and angular momentum. Therefore a Planckon should
have no charge or angular momentum.

Since a black hole can be arbitrarily large - to be able to store
(encode) the information about the original state of a black hole
with a Planckon, infinite different species of Planckons must
exist (Planckons have infinite degeneracy). Such a large reservoir
of quantum states implies huge entropy, which is much larger than
the usual black hole entropy $S_{bh}=\frac{A}{4}$. The excess
entropy may be expressed by the integration constant usually
omitted in the derivation of the black-hole entropy from
$dS=\frac{dE}{T}$. With the presence of an integration constant C
the black hole entropy becomes:
 \be S_{bh}=\frac{A}{4}+C \ee\\
 If one considers an infinite (or very large) C the last stage of the black hole evaporation should be modified by allowing the black hole to tunnel
into a Planckon (i.e a remnant with infinite degeneracy).\\

The main argument raised against the Planckon paradigm is that
having an infinite number of Planckons with, approximately, the
same mass will lead to a divergence in any process with energy
higher than $M_{pl} $. Since there must be a tiny, non zero,
amplitude of Planckon production and since one must sum over all
possible (infinite) species of Planckons, one ends up with an
infinite production rate, which will cause the universe to be
unstable to instantaneous decay into remnants (which is evidently
not the case).

Such an infinite production rate will also plague the coupling of
Planckons to soft quanta ($wavelength \gg l_{pl} )$, where
Planckons can be described by an effective theory, in which the
Planckons are described by a point-like object. Thus the coupling
to soft gravitons, for example, will depend only on its mass and
not on its internal structure or information content and again the
infinite number of species will cause the luminosity to be
infinite and a divergence in the graviton propagator will occur.
Such divergences should have great impact on low energy
physics (due to the coupling between soft quanta and Planckons).\\

Another argument raised against the Planckon is that a planck
scale remnant cannot hold the required information because of
entropy bounds relations between entropy and energy \cite{beken}.

A physical model for a Planckon should deal with the above
problems without invoking any new and unfamiliar physics. A
discussion on the way the model deals with the above problems can
be found in \ref{sec:discussion}.

For further discussion see the excellent reviews
~\cite{preskill,page,giddings,hawk2,frolov,acn} and references
therein.

\section{Models For The Planckon}
\label{sec:models}

This section contains a development of a consistent physical
models for a Planckon using semi-classical methods (i.e. Quantum
fields on curved space-time, WKB approximation etc.).

Such a model will include the main features of the Planckon as
mentioned in \ref{subsec:remnant} (neutral particle with infinite
degeneracy) and will also provide an effective description as to
how the Planckon avoids the estimate of infinite production rates,
despite its inherent infinite degeneracy.

The model is based on \cite{acn}, where the basic properties of
the Planckon were outlined and on \cite{casher1,casher2}, where a
precursor model (that of the "Achronon") is outlined and the
possibility of the existence of the Planckon is briefly discussed.

\subsection{General Properties Of The Models}
\label{subsec:properties}

As the evaporating black hole approaches the Planck scale and the
Compton wavelength of the remaining black hole exceeds its
Schwarzschild radius, quantum effects (especially the uncertainty
principle) become important and a quantum treatment of the system
is required.

From now on, unless stated otherwise, the natural units will be
used $c = G = \hbar = 1$ and the signature is of the form $\left(
{ + , - , - , - } \right)$.

In the models suggested the black hole (whose mass is
approximately $m_{pl} )$ tunnels into a specific state, in which
the mass is distributed at a distance $\Delta \ll 1$ from its
Schwarzschild horizon.

The proper physical description of the Planckon is a soliton with
mass of the order of $m_{pl}$, but, since the knowledge of the
quantum treatment of a soliton is limited to a pertubative
treatment, i.e., expanding the corrections in orders of
$\frac{1}{M_{soliton}}$. At the planck scale the perturbative
 expansion breaks down, since $\frac{1}{M_{soliton}} \backsim
\frac{1}{m_{pl}}\backsim 1$ and a different treatment is required.
In the models suggested, the soliton problem was avoided by
describing the mass configuration of the Planckon with a
spherical-symmetric scalar field. The scalar field generates the
classical geometry (metric). The quantum corrections to the
planckon are given by quantum fields propagating over the
classical geometry. For simplicity only massless quantum scalar
and fermion fields \footnote {The need for super-symmetry will be
explained within the context of the models} are considered. The
mass configurations are chosen specifically to produce a tiny
(almost zero) $g_{00}$ . Looking at the field equations one can
easily see that the time dependence of a field is proportional to
some power of $g_{00}$, which means that the fields are almost
static (this time independence is a manifestation of the
gravitational time dilation). This $g_{00}$ time dependence also
appears in the Einstein field equations, i.e., the metric is
almost static. Since the time dependence of each field is
proportional to $g_{00}$ each vertex will carry a power of
$g_{00}$ and quantum corrections to the soliton will take the form
of a parturbative expansion in $g_{00}$.

The classical scalar field satisfies the following conditions:

\be
\partial _t g_{\mu \nu } = 0\,\,;\,\, - g^{rr} = 1 -
\frac{2M(r)}{r}\,\,;\,\,\partial _t \phi = 0\,\,;\,\,\partial
_\varphi \phi = 0\,\,;\,\,\partial _\theta \phi = 0 \ee

These conditions ensure that the metric and the scalar fields are
spherically symmetric and are "frozen" in time as described above.
The uncertainty principle together with the $g_{00}$ time
dependence will prevent the mass configuration from collapsing due
to gravitational force.\\

Note that this type of configuration cannot be produced by
collapse from infinity and can be reached only by tunneling
\cite{casher1,casher2}.

The expression for the energy-momentum tensor of a scalar field
is:

\be
T_\nu ^\mu = g^{\mu \lambda }\partial _\lambda \phi \partial
_\nu \phi - \frac{1}{2}g^{\mu \lambda }g_{\nu \lambda } g^{\rho
\sigma }\partial _\rho \phi \partial _\sigma \phi
\ee

The above particular conditions generate an energy-momentum tensor
with the following property:

\be
 T_r^r = - T_t^t
\ee

The $T_t^t $ term will be given by the mass density distribution of the
specific model and the metric will be given by the Schwarzschild solution
for the given energy-momentum tensor:

\be
{
 { -g^{rr} = 1 - \frac{2M(r)}{r}} \,\,;\,\,  g_{tt} = \left( {1 -
\frac{2M(r)}{r}} \right) \exp \left( { - 8\pi  \int_{r}^{\infty}{
r'^2\frac{2T^{t}_{t}}{r'g_{rr}}}\, dr'} \right)}
\\
\ee

The effective potentials quantum scalar and fermion fields
(massless S-wave fields) experience, when propagating over curved
background, are:

\be \label{eq:spotential}
 V_{scalar} (r) = \frac{1}{2r}\partial _r \left( {\left( { - g^{11}}
\right)g_{00} } \right) \\
\ee
 \be \label{eq:fpotential}
 V_{fermion} = W^2(r)\pm \frac{\partial W(r)}{\partial \rho }
\approx \frac{k^2}{r^2}g_{00} \\
\ee

Where in the fermion case $W(r) = \frac{\left| k \right|}{r}\sqrt
{g_{00} } $ is a super symmetric potential and thus only the
$W^2(r) = \frac{k^2}{r^2}g_{00} $ term contributes [10, 11].

It is shown, that for a given quantum field, there is a nearly
infinite number of possible quantum excitations.

In this paper the physical properties of two possible models,
which allow the Planckon to have infinite degeneracy but finite
production rate and are also singularity free, are investigated.

\section {Potential Well Model For The Planckon}
\label{sec:well}

In this model the planckon has a total mass M which is located at
$ r=2M+\Delta$ (a distance $\Delta$ from the Schwarzschild horizon
$r=2M$).\\
 The mass distribution is:
\be
m(r)=M \cdot \Theta(r-2M-\Delta)
\\
\ee

This mass distribution gives the following energy distribution:

\be \label{eq:t00}T^0_0=\frac{1}{4\pi r^2}\partial _r M(r)=\frac
{M}{4\pi r^2}\delta (r-2M-\Delta)
\\
\ee

The metric generated by this energy momentum tensor is:

\be \label{eq:metric}
 g_{00}=\left \{ \begin{array}{ll}
 e^{-\frac{4M}{\Delta}} & r\leqslant 2M+\Delta \\
 1-\frac{2M}{r} & 2M+\Delta<r\\
 \end{array} \right.
 \,\,;\,\,
 g_{11}=\left \{ \begin{array}{ll}
 1 & r\leqslant 2M+\Delta \\
- \left (1-\frac{2M}{r} \right)^{-1} & 2M+\Delta<r\\
 \end{array} \right.
\ee

as can be seen the model contains no singularities and no
horizons.

The quantum corrections to the Planckon are given by the quantum
fields propagating in the volume trapped by the classical mass
configuration.

The model is described by first evaluating the energy correction
due to a scalar field and than expanding the model to include the
super-symmetric fermionic partner, in order to cancel the divergence in the self energy.\\

\subsection {Self Energy Of A Scalar Field}
\label{subsec:scalar energy}

The scalar field equation of motion will be of the form:

\be
\partial _\rho ^2 h(r) - \omega ^2h(r) + \frac{1}{2r}\partial _r ( - g^{11}g_{00}
)h(r) = \partial _\rho ^2 h(r) - \omega ^2h(r) = 0 \ee

Where the following definitions are used:

\be \rho = \sqrt { - g_{11} g^{00}} r = \sqrt {g^{00}} r =
re^{\frac{2M}{\Delta }} \ee

The potential is located at:

\be \rho (2{\rm M} + \Delta ) = (2{\rm M} + \Delta
)e^{\frac{2M}{\Delta }}. \ee

Assuming the boundary conditions:

\be h(0) = h(\rho (2{\rm M} + \Delta )) = 0 \ee

The energy eigenvalues are the same as for a potential well.
 The self energy will be of the form:

\be E = \frac{1}{2}\sum \omega = \frac{1}{2}\sum {\frac{n\pi
}{\rho (2{\rm M} + \Delta )}} = \frac{1}{2}\sum\limits_{n =
1}^{\frac{\rho (2{\rm M} + \Delta )}{\pi }\sqrt {V(2{\rm M} +
\Delta )}} {\frac{n\pi }{\rho (2{\rm M} + \Delta )}}\propto \rho
(2{\rm M} + \Delta ) \ee

The self energy ``diverges'' as $\rho (2{\rm M} + \Delta ) =
(2{\rm M} + \Delta )e^{\frac{2M}{\Delta }}$.

To make this model finite one  has to invoke super-symmetry. We
emphasize that exact super-symmetry is not required, but only that
there is an equal number of bosonic and ferminionic degrees of
freedom (need not have the same energy levels). The model is
modified  only by the minimal modifications needed to make it
super-symmetric, which means replacing the scalar field with a
complex scalar field (the self energy will grow by a factor of 2)
and adding a weyl fermion (the super-symmetric partner of the
complex scalar). The total self energy will be the
sum of the two contributions of the self energies of the fields.\\

\subsection{Self Energy Of A Fermion In The Potential Well Model} \
\label{subsec:fermion energy}

The fermion self energy in a spherical potential well will be
derived by following the derivation of the solution for the MIT
bag model in \cite{hecht} and \cite{bag1,bag2}.

The massless Dirac equation is:

\be \gamma^\mu\nabla_\mu \psi = 0 \ee

where $\nabla_{\mu}=\partial_{\mu}-\Gamma_{\mu}$ and
$\Gamma_{\mu}$ is the spin-connection in the vierbeins method.

The solutions for $\psi $ are of the form:

\be \label{eq:fsol} {\begin{array}{*{20}c}
 {\psi = \left( {\begin{array}{l}
 g(r)Y_{[l_a \frac{1}{2}]jm_j } \\
 \pm f(r)Y_{[l_b \frac{1}{2}]jm_j } \\
 \end{array}} \right)}\\
\end{array} }
\ee

the $\pm$ is defined for:

 \be k = \mp (j + \frac{1}{2}) = \left\{
{{\begin{array}{*{20}c}
 {\begin{array}{l}
 - \left( {j + \frac{1}{2}} \right) = - (l_a + 1) \\
 \left( {j + \frac{1}{2}} \right) = l_a \\
 \end{array}} \hfill & {\begin{array}{l}
 j = l + \frac{1}{2} \\
 j = l - \frac{1}{2} \\
 \end{array}} \hfill \\
\end{array} }} \right.
\ee

 Where the following definitions have been used:
\be
 j = (l_a + \frac{1}{2})\,\,;\,\,l_b = (l_a + 1)\,\,for\,\,k < 0 \\
 \ee
 \be
 j = (l_a - \frac{1}{2})\,\,;\,\,l_b = (l_a - 1)\,\,for\,\,k > 0 \\
\ee

k is the dirac quantum number which differentiates the two states
of opposite parity for each value of j.\\

If one defines $G(r) = r \cdot g(r)$and $F(r) = r \cdot f(r)$ one
gets the following equations:

\be
 \left( { - \sqrt {\left( { - g^{11}} \right)g_{00} } \partial _r - \sqrt
{g_{00} } \frac{k}{r}} \right)G + \omega F = 0 \\
\ee
\be
 \left( {\sqrt {\left( { - g^{11}} \right)g_{00} } \partial _r - \sqrt
{g_{00} } \frac{k}{r}} \right)F + \omega G = 0 \\
\ee

Substituting the last 2 equations into each other one gets:

\be \partial _{\rho }'^2 F + \left( {\omega ^2 - \frac{k(k -
1)}{{\rho }'^2}}
\right)F = 0 \\
 \ee
\be
\partial _{\rho }'^2 G + \left( {\omega ^2 - \frac{k(k + 1)}{{\rho }'^2}}
\right)G = 0 \ee\\

Where $\rho$ has the is the same as in (24) (the same
as for the scalar field).\\

Since ${\begin{array}{*{20}c}
 {j = (l_a \pm \frac{1}{2})} \hfill & {k = \mp } \hfill \\
\end{array} }\left( {j + \frac{1}{2}} \right)$ one has:

\be
 k = - (l_a + 1) \Rightarrow \left\{ {\begin{array}{l}
 k(k + 1) = l_a (l_a + 1) \\
 k(k - 1) = (l_a + 1)(l_a + 2) = l_b (l_b + 1) \\
 \end{array}} \right. \\
\ee
 \be
 k = l_a \Rightarrow \left\{ {\begin{array}{l}
 k(k + 1) = l_a (l_a + 1) \\
 k(k - 1) = l_a (l_a - 1) = l_b (l_b + 1) \\
 \end{array}} \right. \\
\ee

If one defines $\rho = \omega {\rho }'$, the equations will take
the general form of:

\[
{\begin{array}{*{20}c}
 {\left( {\frac{\partial }{\partial \rho ^2} - \frac{l(l + 1)}{\rho ^2} + 1}
\right)u_l = 0} \hfill & {u_{l_a } = G\,\,;} \hfill \\
\end{array} }\,\,u_{l_b } = F
\]

Where $u_l $ is the solution for the one dimensional radial
equation - in this case $u_l $ is the spherical Bessel function
and since one only considers solutions, which are regular at the
origin, one has to choose:

\[
\frac{G(\rho )}{\rho } = j_{l_a } (\rho )\,\,;\,\,\frac{F(\rho
)}{\rho } = \mp j_{l_b } (\rho )
\]

Where the sign in $F(\rho )$ are for $ - (l_a + 1)$ and the + sign
is for $k = l_a $.\\

Incorporating the results into (\ref{eq:fsol}) one obtains:

\be
\psi = N\left( {\begin{array}{l}
 j_{l_a } (\omega \rho )Y_{[l_a \frac{1}{2}]jm_j } \\
 - j_{l_b } (\omega \rho )Y_{[l_b \frac{1}{2}]jm_j } \\
 \end{array}} \right)
\ee
Where N is a normalization factor.\\

Our boundary conditions, to ensure the confinement of the fermion
field inside $r = 2{\rm M} + \Delta $, are:

\begin{enumerate}
 \item  \begin{center}
 ${\begin{array}{*{20}c}
 {\int\!\!\!\int {\bar {\psi }\left( {\vec {\gamma } \cdot \frac{\vec
{r}}{r}} \right)\psi r^2d\Omega } } \hfill & {r = 2{\rm M} +
\Delta } \hfill
\\
\end{array} }$
\end{center}

 \item
 \begin{center}
  ${\begin{array}{*{20}c}
 {\int\!\!\!\int {\bar {\psi }\psi r^2d\Omega = 0} } \hfill & {r = 2{\rm M}
+ \Delta } \hfill \\
\end{array} }$
\end{center}
\end{enumerate}

The first condition ensures that there is no probability density
current in the radial direction from the sphere of radius $r =
2{\rm M} + \Delta $, while the second condition ensures that the
Lorentz scalar quantity $\bar {\psi }\psi $ (the probability of
finding the particle) will be zero over a sphere with radius $r =
2{\rm M} + \Delta $.\\
The first condition is satisfied automatically from the
orthonormality of the spherical harmonics.

Putting the solutions into the boundary condition and defining $a
= \left( {2{\rm M} + \Delta } \right)e^{\frac{2M}{\Delta }}$ the
second condition becomes:

\be \label{eq:second condition}
\begin{array}{l}
 \int\!\!\!\int {a^2\left( {\left( {j_{l_a } \left( {\omega a} \right)}
\right)^2\left| {Y_{[l_a \frac{1}{2}]jm_j } } \right|^2 - \left(
{j_{l_b } \left( {\omega a} \right)} \right)^2\left| {Y_{[l_b
\frac{1}{2}]jm_j } }
\right|^2} \right)d\Omega } \\
 = \left( {\left( {j_{l_a } \left( {\omega a} \right)} \right)^2 - \left(
{j_{l_b } \left( {\omega a} \right)} \right)^2} \right) = 0 \\
 \end{array}
\ee

equation \eqref{eq:second condition} is satisfied if:

\[
j_{l_a } (\omega a) = \frac{k}{\left| k \right|}j_{l_b } (\omega
a)
\]

Let us denote the $n^{th}$ solution of this equation for a
specific value of k as $\chi _{n,k} = \omega _{n,k} a$ and so the
energies will be of the form:

\[
\omega _{n,k} = \frac{\chi _{n,k} }{a}
\]

Since only orbital momentum l=0 is considered, the interest is
limited to the case where $k = - 1\mathop (j = \frac{1}{2},l_a =
0,l_b = 1)$, which gives the equation:

\[
j_0 (\chi _{n, - 1} ) = j_1 (\chi _{n, - 1} )
\]

The last equation can be written as follows:

\be \label{eq:energy equation}
\tan \left( {\chi _{n, - 1} }
\right) = - \frac{\chi _{n, - 1} }{\chi _{n, - 1} - 1} \ee

The first few numerical solutions to \eqref{eq:energy equation}
are:
\[
 \chi _{1, - 1}  = 2.0427869 \quad \chi _{2, - 1} = 5.396016118 \quad \chi _{3, - 1} = 8.577558785 \quad \chi _{4, - 1} = 11.73650396
\]

One can easily show that the distances between the solutions
converge rapidly (from above) to $n\pi $, so one can approximate
the energy levels by (a lower bound):

\[
E_f = \sum\limits_{n = 1}^{\frac{a}{\pi }\sqrt{V(2{\rm M} + \Delta
)}} {\frac{\chi _{n, - 1} }{a}} \approx \sum\limits_{n =
1}^{\frac{a}{\pi }\sqrt {V(2{\rm M} + \Delta )}} {\frac{2.043 + (n
- 1)\pi }{a}} = \sum\limits_{n = 1}^{\frac{\rho (2{\rm M} + \Delta
)}{\pi }\sqrt{V(2{\rm M} + \Delta )}} {\frac{2.043 + (n - 1)\pi
}{\rho \left( {2{\rm M} + \Delta } \right)}}
\]

\subsection {Total energy of the super-symmetric potential well
model} \label{subsec:total energy spherical well}

The total self-energy (bounded from above) of the model will be:

\be \label{eq:total self-energy spherical well}
\begin{array}{l}
 E_{self} = E_s - E_f < \sum\limits_{n = 1}^{\frac{\rho (2{\rm M} + \Delta )
\sqrt{V(2{\rm M} + \Delta )}}{\pi }} {\frac{n\pi }{\rho (2{\rm M}
+ \Delta )}} - \sum\limits_{n = 1}^{\frac{\rho (2{\rm M} + \Delta
)\sqrt{V(2{\rm M} + \Delta )}}{\pi }} {\frac{2.043 + (n - 1)\pi
}{\rho \left( {2{\rm M} + \Delta }
\right)}} \\
\\
 = \sum\limits_{n = 1}^{\frac{\rho (2{\rm M} + \Delta )\sqrt{V(2{\rm M} + \Delta
)}}{\pi }} {\frac{1.1}{\rho (2{\rm M} + \Delta )}} =
\frac{1.1}{\pi }\sqrt{V(2{\rm M} + \Delta )} =
0.35\sqrt{\frac{\Delta }{\left( {2M +
\Delta } \right)^3}} \\
 \end{array}
\ee

Using ${V_{scalar}}(2{\rm M} + \Delta)\approx{V_{fermion}}(2{\rm
M} + \Delta)=V(2{\rm M} + \Delta)$, which is correct if $\Delta
\ll \rm M$.

The total energy will be of the form:

\be \label{eq:total energy spherical well} E = \sqrt{{\rm M}^2 +
\frac{Const}{2{\rm M}(2{\rm M} + \Delta )^2}} +
0.35\sqrt{\frac{\Delta }{\left( {2M + \Delta } \right)^3}} \ee
\\

The $M^2$ term is the rest energy of the planckon, the
$\frac{Const}{2{\rm M}(2{\rm M} + \Delta )^2}$ term is the kinetic
energy due to the uncertainty principle and the $
0.35\sqrt{\frac{\Delta }{\left( {2M + \Delta } \right)^3}}$ is the
quantum self energy of the planckon.

 Note that the contribution to the self energy comes
mainly from trans-planckian modes i.e. modes with wavelength,
which is lower than the Planck length.

\subsection {Discussion On The Self Energy}
\label{subsec: discussion self energy}

The self energy expression \eqref{eq:total self-energy spherical
well} will now be examined. Expression \eqref{eq:total self-energy
spherical well} is not only finite but also small (since
$\Delta\ll\rm M$). This is of high importance since, otherwise,
the black hole would not have tunnelled into the planckon due to
energy conservation.
 Expression \eqref{eq:total self-energy spherical well} was calculated for
only 2 super-partner fields out of the total number of fields
(this number should  be about several hundreds and will denoted as
$C_f$). For each super-multiplet the self energy expression should
be proportional to \eqref{eq:total self-energy spherical well} so
the correct expression for the self energy has the form:

\be \label{eq:corrected self-energ} E_{self}=k \cdot C_f
\sqrt{\frac{\Delta }{\left( {2M + \Delta } \right)^3}}
 \ee\\

Where k is an unknown factor (which might also be negative) due to
the contribution of each of the super-multiplets and $C_f$ is the
factor due to the total number of fields (or super-multipets). The
expression \eqref{eq:corrected self-energ} is also small if one
demands that:
 \be
 \Delta < \frac{(2M)^3} {(k \cdot C_f)^2}
\ee \\
The last expression gives some limitations of the value of
$\Delta$, which was arbitrary up to now. The finiteness of the
self energy is an important example for a scenario, in which, the
infinite degeneracy of the planckon (which is proportional to the
number of energy levels) does not give a divergent expression but
a small finite expression because of the coupling (proportional to
$g_{00}$). The same effect should happen in each order of quantum
loop corrections in field theory hence rendering the effect of the
infinite degeneracy of the planckon finite and no divergence will
occur. It should be noted that super-symmetry had to be
incorporated to achieve finite self-energy in the above example.

\subsection{Physical Properties Of The Super-Symmetric Potential
Well Model}
 \label{subsec:physical properties spherical well}

As mentioned above this model has no horizon and no singularity
(as expected from a quantum gravity theory).

The mass of the Planckon is approximately $m_{pl} $ since the self
energy is much smaller than the classical mass and the energy term
that comes from the uncertainty effect.

There are approximately $g^{00} = e^{\frac{4M}{\Delta }}$ possible
states, which are effectively degenerate, since the separation
between the states is in the order of $\sqrt {g_{00} }$.

The only way for a field to interact with the internal degrees of
freedom is to enter the volume inside the mass. The time for such
an interaction to take place, for any observer who observes the
interaction from outside the Planckon, is of the order of $ \sim
\frac{1}{\sqrt {g_{00}}} = e^{\frac{2M}{\Delta }}$, because of the
gravitational time dilation due to the inner metric.

Effects, such as the time dilation, can make the production rate
of a Planckon effectively zero (despite the Planckon's huge
degeneracy), by rendering the Planckon coupling much smaller than
the degeneracy. An example for such a scenario was presented by
the self energy calculations in \ref{sec:well} and discussed in
\ref{subsec: discussion self energy}.

As a consequence of the Planckon vanishing coupling the only
possibility to experimentally find evidence for a Planckon is
through its gravitational effects since it has no charge (charges
are emitted through tunnelling).

The main drawback of the spherical well model is the
$\delta$-function divergence of the energy-momentum tensor. The
following model does not suffer from the same problem but its self
energy "diverges" as $\sqrt{g^{00}}$.

\section{A Linear Model For The Planckon}
\label{sec:linear model}

In this model the total mass M is distributed linearly in the
following way:

\[
 m(r)=\left \{{\begin{array}{ll}
 \frac {r-\Delta}{2} & 0<r \leq {2{\rm M+\Delta}}\\
  {\rm M} & \textrm \quad {else}\\
 \end{array}} \right.
 \]

The energy momentum tensor behave as:

\[
T^t_t(r)=\left \{{\begin{array}{ll}
 \frac {1} {8\pi r^2} & 0<r \leq {2{\rm M+\Delta}}\\
 0 & \textrm \quad {else}\\
 \end{array}} \right.
\]

The metric has the form:

\[
{\begin{array}{ll}
  - g^{rr} = \left \{{\begin{array}{ll}
   1 & r \leq \Delta\\
   \frac {\Delta}{r} & \Delta < r < 2{\rm M}+\Delta\\
   1-\frac{2\rm M}{r} & 2{\rm M}+\Delta \leq r\\
   \end{array}}\right.
   & g_{00}= \left \{{\begin{array}{ll}
   e^{ - \frac{4{\rm M}}{\Delta }} & r \leq \Delta\\
   \frac {\Delta}{r}e^{-\frac{2}{\Delta}(2{\rm M}+\Delta-r)} & \Delta < r < 2{\rm M}+\Delta\\
   1-\frac{2\rm M}{r} & 2{\rm M}+\Delta \leq r\\
   \end{array}}\right.
   \end{array}}
   \]\\

\subsection {The Self Energy Of A Complex Scalar Field In The
Linear Model} \label{subsec:linearscalar}

The effective potential a scalar field experiences due to the
curved background is:

\[
V_{scalar} (r) = \frac{1}{2r}\partial _r \left( {\left( { -
g^{11}} \right)g_{00} } \right) = \left( {\Delta r^{ - 3} - \Delta
^2r^{ - 4}} \right)e^{ - \frac{2}{\Delta }\left( {2{\rm M} +
\Delta - r} \right)} \qquad \Delta\leq r \leq2M+\Delta\]

To calculate the energy of the complex scalar field the WKB
approximation has been used.

The Bohr-Sommerfeld quantization condition (up to some constant in
the left hand side) is given by :

\[
n\pi = \int_{0}^{\rho '} {\sqrt {\omega ^2 - V({\rho }')} d{\rho
}' = } \frac{1}{\sqrt \Delta }\int\limits_\Delta ^r {{r}'\sqrt
{\left( {r^{ - 3} - \Delta r^{ - 4}} \right)e^{\frac{2}{\Delta }(r
- {r}')} - \left( {{r}'^{ - 3} - \Delta {r}'^{ - 4}} \right)}
d{r}'}
\]

From the quantization condition one gets the density of states:

\[
\frac{dn}{dr} = \frac{\left( {2r^2 - 5\Delta r + 4\Delta ^2}
\right)}{2\pi \Delta ^{\frac{3}{2}}r^3}\int\limits_\Delta ^r
{\frac{{r}'^3e^{\frac{2}{\Delta }(r - {r}')}}{\sqrt {\left( {r -
\Delta } \right){r}'^4e^{\frac{2}{\Delta }(r - {r}')} - \left(
{{r}' - \Delta } \right)r^4} }d{r}'}
\]

The complex scalar self-energy is given by:

\[
\begin{array}{l}
 \left\langle E \right\rangle _{scalar} = 2 \cdot \frac{1}{2}\sum {\omega
\approx \sum\limits_{n = 1}^{n_{\max } } {\sqrt {V(r)} \simeq
\int\limits_1^{n_{\max } } {\sqrt {V(r)} dn} = \int\limits_\Delta
^{2{\rm
M} + \Delta } {\sqrt {V(r)} \frac{dn}{dr}dr} } } \\
 = \int\limits_\Delta ^{2{\rm M} + \Delta } {\frac{\sqrt {\left( {r -
\Delta } \right)} \left( {2r^2 - 5\Delta r + 4\Delta ^2}
\right)e^{ - \frac{1}{\Delta }\left( {2{\rm M} + \Delta - r}
\right)}}{2\pi \Delta r^5}\left( {\int\limits_\Delta ^r
{\frac{{r}'^3e^{\frac{2}{\Delta }(r - {r}')}}{\sqrt {\left( {r -
\Delta } \right){r}'^4e^{\frac{2}{\Delta }(r -
{r}')} - \left( {{r}' - \Delta } \right)r^4} }d{r}'} } \right)dr} \\
 \end{array}
\]

This expression diverges as $\sqrt{g^{00} }  = e^{\frac{2M}{\Delta
}}$ and again, in an attempt to obtain a finite expression,
super-symmetry is invoked in the same way as in the spherical well
model (adding the fermion super-partner of the complex scalar).\\

\subsection {The Self Energy Of A Fermion In The Linear Model}
\label{subsec:mylabel7}

The effective potential the fermion experiences, due to the curved
background is:

\[
V_{fermion} = W^2(\rho ) = \frac{k^2}{r^2}g_{00} = \frac{\Delta
}{r^3}e^{ - \frac{2}{\Delta }\left( {2{\rm M} + \Delta - r}
\right)} \qquad \Delta\leq r \leq2M+\Delta \]

To calculate the energy of the fermion field the WKB
approximation has been used in the same way as for the complex scalar.\\
 From the bohr-sommerfeld quantization condition one gets the density of
states:

\[
\frac{dn}{dr} = \frac{2r - 3\Delta }{2\pi \Delta
^{\frac{3}{2}}r^{\frac{5}{2}}}\int\limits_\Delta ^r
{\frac{{r}'^{\frac{3}{2}}e^{\frac{2}{\Delta }(r - {r}')}}{\sqrt
{{r}'^3e^{\frac{2}{\Delta }(r - {r}')} - r^3} }d{r}'}
\]

The total self energy of a fermion is:

\[
\begin{array}{l}
 \left\langle E \right\rangle = 2 \cdot \frac{1}{2}\sum {\omega \approx
\sum\limits_{n = 1}^{n_{\max } } {\sqrt {V(r)} \simeq
\int\limits_1^{n_{\max } } {\sqrt {V(r)} dn} = \int\limits_\Delta
^{2{\rm M} + \Delta } {\sqrt
{V(r)} \frac{dn}{dr}dr} } } \\
 = \int\limits_\Delta ^{2{\rm M} + \Delta } {\frac{\left( {2r^2 - 3\Delta
r} \right)e^{ - \frac{1}{\Delta }\left( {2{\rm M} + \Delta - r}
\right)}}{2\pi \Delta r^4}\left( {\int\limits_\Delta ^r
{\frac{{r}'^{\frac{3}{2}}e^{\frac{2}{\Delta }(r - {r}')}}{\sqrt
{{r}'^3e^{\frac{2}{\Delta }(r - {r}')} - r^3} }d{r}'} } \right)dr} \\
 \end{array}
\]

This expression's divergent behavior is the same as for the
complex scalar field.

\subsection {The Total Self Energy Of The Super-Symmetric Linear
Model}\label{subsec:mylabel7}

The total self energy for the super-symmetric linear model is
given by:

\[\begin{array}{l}
 \left\langle E \right\rangle = \left\langle E \right\rangle _{boson} -
\left\langle E \right\rangle _{fermion} \nonumber\\
\mathrel{\mathop{\kern0pt\longrightarrow}\limits_{\Delta \ll r}}
\int\limits_\Delta ^{2{\rm M} + \Delta } {\frac{\left( {\left(
{2r^2 - 5\Delta r} \right) - \left( {2r^2 - 3\Delta r} \right)}
\right)e^{ - \frac{1}{\Delta }\left( {2{\rm M} + \Delta - r}
\right)}}{2\pi \Delta r^5}\left( {\int\limits_\Delta ^r
{\frac{{r}'^{\frac{3}{2}}e^{\frac{2}{\Delta }(r - {r}')}}{\sqrt
{{r}'^3e^{\frac{2}{\Delta }(r - {r}')} - r^3} }d{r}'} } \right)dr} \\
 \sim - \int\limits_\Delta ^{2{\rm M} + \Delta } {\frac{e^{ -
\frac{1}{\Delta }\left( {2{\rm M} + \Delta - r} \right)}}{\pi
r^4}\left( {\int\limits_\Delta ^r
{\frac{{r}'^{\frac{3}{2}}e^{\frac{2}{\Delta }(r - {r}')}}{\sqrt
{{r}'^3e^{\frac{2}{\Delta }(r - {r}')} - r^3} }d{r}'} }
\right)dr} \\
 \end{array}
 \]
 The energy is still divergent but the divergence
is smaller by a factor of $ \sim \frac{\Delta }{r}$, which is
obviously not enough since the divergence is exponential.

The reason that the divergence is not totally eliminated is that
the effective potential of the scalar and the fermion are the same
only to the first order in $\frac{\Delta }{r}$.

\subsection {Methods Of Reducing The Divergence}
\label{subsec:methods of reducing}

Several methods have been examined in order to reduce the
divergence of the self-energy such as finding different geometries
that will allow one to have some other parameters, besides $\Delta
$, to control the divergences. However as long as one keeps a
linear section in the mass distribution, one ends up with similar
divergences. Some attempts to reduce the divergence were to insert
other consideration such as tunnelling \footnote{The divergence in
the self energy expression may be eliminated by imposing a cutoff
at r=M. The only reason to impose such a cutoff is due to
tunnelling effect} and measurement theory considerations but they
are not directly connected to the self energy and as such can not
help to solve the problem. \\
Another possibility for divergence reduction, which was not
considered, is including in the self energy computation the whole
gravity super-multiplet i.e. the gravitino related vacuum
diagrams, which were not included in the computation.

\subsection {Physical Properties Of The Super-Symmetric Linear
Model}\label{subsec:mylabel9}

The linear model has all the physical properties of the spherical
well model (see section \ref{subsec:physical properties spherical
well}). This model is also physical, since the metric is
continuous and the energy-momentum tensor is not a delta-function,
but a finite regular function.

Currently, the main drawback of the linear model is the self
energy divergence. If not eliminated (or at least shown to be
reduced by other vacuum diagrams that were not taken into account)
this divergence will prevent the black hole from tunneling into
the Planckon, due to energy conservation. The fact that the self
energy of the spherical well model is finite, raises the hope,
that a method can be found which will render the self-energy of
the linear model finite (perhaps by considering contributions from
the gravity super-multiplet as was mentioned in
\ref{subsec:methods of
reducing}).\\

%%\newpage

\section{Discussion}
\label{sec:discussion}

It has been shown, that models exist, which have the general
properties needed to make the planckon physically possible. These
models render most of the arguments against the planckon
non-relevant, especially the loop divergence
arguments\footnote{There is also the argument mentioned in
\ref{subsec:remnant} about the coupling of a soft graviton to a
planckon anti-planckon, but this argument have no physical ground,
since a soft planckon will not be able to create a planckon
anti-planckon pair, because of energy conservation and the
extrapolation of the interactions of gravitons from low-energy
physics into planck scale energy physics is not valid, since there
is possibly a new and different physics at the planck scale}.\\
The other argument mentioned in \ref{subsec:remnant} is based on
entropy bounds, which were derived using adiabatic processes such
as lowering a box into a black hole. Arguments based on adiabatic
continuous processes are irrelevant for the Planckon, as described
here, for two main reasons:
\begin{enumerate}
 \item[-] In general, entropy bounds only measure the difference of
entropy of the systems caused by the process and not the initial
entropy of the systems. The Planckon can hold a huge amount of
information, while being envolved in processes that change the
total entropy of the whole system by a small amount.

\item[-] The specific models of the Planckon presented here and in
\cite{acn,casher1,casher2} , where the process of a creation of a
Planckon contains quantum processes such as tunnelling, cannot be
described by continuous adiabatic processes such as the ones used
for deriving the entropy bounds.
\end{enumerate}

Another argument, which can be raised against the models described
here, is that the Planckon should have infinite degeneracy (not
just very high degeneracy). The argument goes as follows:\\
 A black hole can swllaow a Planckon. If a Planckon is the final state of a black hole it should keep the
information of the black hole and the swallowed Planckon. The only
way to achieve that goal is by requiring the Planckon to have infinite degeneracy.\\
A possible resolution can be obtained by considering black holes
having an internal \footnote{The meaning of internal is that it
does not affect the metric generated by the black hole so that
black hole theory will not have to be changed} Planckon counter.
The value of the counter is the number of swallowed Planckons. A
black hole whose counter has the value N evaporates into N+1
Planckons when its energy reaches N+1 times the Planckon mass.
This allows the Planckon to have very high yet finite degeneracy.

 One can see that the models described in this paper solve
the main problems of black hole remnants, although there are still
many open questions:

\begin{enumerate}
\item \textbf{Making the spherical well model physical} - The main
drawback of the spherical well model is the divergence of the
energy-momentum tensor, since it has the form of a delta function
(which in turn creates a discontinuity in the metric). Possible
extensions to the given model may include extensions adjustments
that will make the metric continuous by allowing the
energy-momentum tensor to be distributed over a finite
non-vanishing region, while keeping the self energy from diverging
(most likely by small perturbations of the energy-momentum tensor)

\item \textbf{Making the self energy of the linear model finite} -
The main drawback of the linear model is the divergence of its
self energy. Elimination of the self energy might be achieved by
methods described in \ref{subsec:methods of reducing}.

\item \textbf{Determining the value of $\Delta$ and $\rm M$} - The
ratio $\frac{\Delta}{\rm M}$ is of great importance for the given
models, but nowhere in the models are the exact values of neither
$\Delta$ nor $\rm M$ calculated. To calculate the value of $\rm
M$, one needs a dynamical model of the Hawking radiation near the
planck scale. The value of $\Delta$ poses more problems, since the
models do not give any method of determining its size. Also, it
has no apparent scale and its size may be much lower than the
planck scale, raising the question of the minimal length scale in
physics. The most probable way to determine $\Delta$ is by finding
the minimum value of the energy, which as for now is not within
reach, due to the number of different fields involved up to the
planck scale.

\item \textbf{Finding a model with minimal self energy } - Since
 two models were introduced, one of which diverges while the other gives a
finite small result, a variational principle might be used to
claim, that a model with minimal self energy exists. Finding such
a model is closely connected to the problem of determining the
value of $\Delta$ and $\rm M$, since their values and ratio
determine the self energy.

\end{enumerate}

If the Planckon exists it should dominate the planck scale
spectrum. As such the models may provide hints, as to what
properties planck scale fields are expected to have and might help
to shed some light on some of the unsolved problems in quantum
gravity and astrophysics such as the information paradox, dark
matter, cosmological constant and different questions related to
planck scale physics \cite{acn,casher1,casher2}.

\end {document}